\documentclass{emulateapj}

\usepackage{epsfig,graphicx,latexsym,amsmath,amssymb}
\usepackage{natbib}
\bibpunct[,]{(}{)}{;}{a}{}{,}

\shorttitle{Efficient Merger of Binary Supermassive Black Holes in Merging Galaxies}
\shortauthors{F. M. Khan et al.}

\begin{document}


\title{Efficient Merger of Binary Supermassive Black Holes in Merging Galaxies}


\author{FAZEEL MAHMOOD KHAN\altaffilmark{1,2}}
\affil{$^1$Astronomisches Rechen-Institut, Zentrum f\"ur Astronomie, 
Univ. of Heidelberg, M{\"o}nchhof-Strasse 12-14, 69120 Heidelberg, Germany\\
$^2$Department of Physics, Government College University (GCU), 54000  
Lahore, Pakistan\\}

\author{ANDREAS JUST}
\affil{Astronomisches Rechen-Institut, Zentrum f\"ur Astronomie, 
Univ. of Heidelberg, M{\"o}nchhof-Strasse 12-14, 69120 Heidelberg, Germany\\}

\and

\author{DAVID MERRITT}
\affil{Department of Physics and Center for Computational Relativity and Gravitation, 85 Lomb Memorial Drive, Rochester Institute of Technology
Rochester, NY 14623-5604 \\}

\begin{abstract}
In spherical galaxies, binary supermassive black holes (SMBHs) have difficulty reaching sub-parsec separations due to depletion of stars on orbits that intersect the massive binary - the ``final parsec problem.''
Galaxies that form via major mergers are substantially nonspherical, and it has
been argued that the centrophilic orbits in triaxial galaxies might
provide stars to the massive binary at a high enough rate to avoid stalling.
Here we test that idea by carrying out fully self-consistent merger simulations of galaxies containing central SMBHs.
We find hardening rates of the massive binaries that are indeed much higher than
in spherical models, and essentially independent
of the number of particles used in the simulations.
Binary eccentricities remain high throughout the simulations.
Our results constitute a fully stellar-dynamical solution to the final parsec problem
and imply a potentially high rate of events for low-frequency gravitational wave detectors like
LISA.
\end{abstract}



\section{Introduction}\label{sec-intro}
 
The prospect that low-frequency gravitational radiation will be detected
by LISA, the laser-interferometer space antenna 
\citep{Hughes03,BC04},
 has motivated theoretical studies into the formation and
evolution of binary supermassive black holes (SMBHs).  
Such binaries would constitute the highest signal-to-noise ratio 
sources of low-frequency gravitational waves, but as is the 
case for virtually all potential LISA sources, the event rate is still poorly known,
with estimates ranging from a few to a few thousand events per year
\citep[e.g.][]{WL03,RW05, sesana10}.
While astronomical evidence for the existence of SMBHs with masses $M_\bullet>10^6
M_\odot$ in galaxy spheroids beyond the Local Group
is increasingly compelling \citep{FF05},
the evidence for SMBHs with masses $\leq 10^6 M_\odot$,
which lie more nearly in the LISA band, is still largely circumstantial
\citep{vdm04}, and the evidence for binary SMBHs is weaker still
\citep{komossa06}.

When two galaxies merge, a binary SMBH 
forms at the center of the new galaxy \citep{BBR80,roos81}.  
A common practice when estimating event rates for LISA
is to equate the binary SMBH coalescence rate 
with the galaxy merger rate, the latter derived from 
models of structure formation in which galaxies merge hierarchically \citep{Hae94,MHN01,VHM03,WL03,JB03}. 
However it is also possible that coalescence is delayed,
perhaps indefinitely, if the massive binary is limited in its ability 
to exchange angular momentum with surrounding stars and gas.
This is sometimes referred to as the ``final parsec problem''
\citep[e.g.][]{MM03} since at this approximate separation, 
a massive binary in a spherical galaxy would
efficiently eject all stars on intersecting orbits \citep{MF04,ber05}.
Further evolution of the binary is possible, even in the absence of gas, if these 
orbits are replenished, or if the galaxy geometry allows for a much larger
population of centrophilic orbits than in the spherical case \citep{MP2004,ber05}.

In this paper, we argue that the latter situation may be generic.
We carry out direct $N$-body merger simulations of galaxies
containing central SMBHs,
and show that the binary SMBHs formed in such simulations
continue to shrink, via interaction with stars, well beyond the point
at which they would be expected to stall in spherical galaxies.
Of course, this mechanism can act in addition to torques from gas, if present,
\citep[e.g.][]{escala05,dotti07,cuadra09}, or to loss-cone repopulation due to ``massive perturbers'' \citep{perets08}, etc.
However it appears that collisionless stellar dynamics is sufficient.

In \textsection 2 we describe the numerical methods and initial conditions. 
Simulations of isolated (spherical) models are presented in \textsection 3,
while \textsection 4 describes the results of galaxy merger simulations. 
\textsection 5 estimates time scales for coalescence of massive binaries, and
\textsection 6 sums up.

\section{Numerical Methods and Initial Conditions}\label{sec-model}

Our initial conditions are based on spherical galaxy models following 
Dehnen's (1993) density law:
\begin{equation}
\rho(r) = \frac{(3-\gamma)M_\mathrm{gal}}{4\pi}\frac{r_{0}}{r^{\gamma}(r+r_{0})^{4-\gamma}}, \label{denr}\\
\end{equation}
and with cumulative mass profile
\begin{equation}
M(r) = M_\mathrm{gal}\left(\frac{r}{r+r_{0}}\right)^{3-\gamma}, \label{massr} 
\end{equation}
where $M_\mathrm{gal}$ is the total mass of the galaxy, $a$ is a scale radius, and $\gamma$ defines the inner density profile slope. 
We adopted $\gamma=1$, corresponding to a Hernquist model \citep{Hern90}, for all initial models.
A massive particle representing a SMBH was placed at the center of each galaxy model. 
The isotropic distribution function reproducing $\rho(r)$ in the combined potential of the stars and the SMBH in dynamical equilibrium was computed numerically and used to generate Monte-Carlo positions and velocities for the stars.
In what follows, we adopt ``model units'' such that $M_\mathrm{gal} = G = r_{0} = 1$.

The models can be scaled to physical units by fixing galaxy mass $M_\mathrm{gal}$ and scale radius $r_0$ combined with the relations
\begin{subequations}
\begin{eqnarray}
\left[T\right] &=& \left(\frac{GM_\mathrm{gal}}{r_{0}^{3}}\right)^{-1/2}\nonumber\\
    &=& 1.5 \mathrm{Myr} \left(\frac{M_\mathrm{gal}}{10^{11}M_{\sun}}\right)^{-1/2} \left(\frac{r_{0}}{\mathrm{kpc}}\right)^{3/2},\\
\left[V\right] &=& \left(\frac{GM_\mathrm{gal}}{r_{0}}\right)^{1/2}\nonumber\\
    &=& 655 \mathrm{km\,s^{-1}} \left(\frac{M_\mathrm{gal}}{10^{11}M_{\sun}}\right)^{1/2} \left(\frac{r_{0}}{\mathrm{kpc}}\right)^{-1/2}
\end{eqnarray}
\end{subequations}
In a Hernquist model the effective (projected half-mass) radius $r_\mathrm{e}$ is $\sim 1.81r_{0}$. 
For a typical, luminous ($M_\mathrm{B}\approx-20$)
elliptical galaxy or bulge, $r_\mathrm{e}\approx 1.5$\, kpc, 
$M_\mathrm{gal}\approx 10^{11}M_{\sun}$ and $[T]\approx 1.1$\,Myr.
Effective radii scale very weakly with galaxy luminosity (though with considerable
scatter; e.g. \citet{Ferrarese06}) and in the case of $10^9M_\odot$ galaxy, 
whose SMBH would have a mass more suited to detection by LISA, 
the physical unit of time would be closer to 6Myr.
In the latter case, the length of our longest integrations amounts to 
$\sim 1.5$ Gyr.

Two sets of simulations were carried out. In the first set, a single spherical
galaxy was created, with a central SMBH of mass $M_\bullet=0.001$ (Model A) or $0.01$ (Model B).
(In what follows, $M_\bullet$ is always used to denote the mass of a single SMBH.)
A second SMBH, of the same mass, was then placed at a distance of $0.5$ from the center on a circular orbit. 
Integrations were continued until a time $t_\mathrm{max} = 100$.

In the second set, two identical spherical models were created, each containing a single central SMBH, and the two galaxies were merged.   Here again, two values for $M_\bullet$ were used: $M_\bullet=0.001$ (Model C) or $0.01$ (Model D) and the integrations were continued until a time $t_\mathrm{max} = 250$. 

In order to test the dependence of the results on particle number $N$, all simulations were carried out with $N = (32, 64, 128, 256, 512)\times 10^3$. 
Tables \ref{TableAB} and \ref{TableCD} summarize the parameters of the two sets of runs.

\begin{table}
\caption{Parameters of the single-galaxy integrations} 
\begin{tabular}{c c c c c c}
\hline
Run & $N$ & $M_\bullet$ & Run & $N$ & $M_\bullet$ \\
\hline
A1& 32k &	$0.001$	& B1 & 32k  & 0.01\\
A2& 64k &	$0.001$	& B2 & 64k  & 0.01\\
A3 & 128k &     $0.001$ & B3 & 128k & 0.01 \\
A4 & 256k &     $0.001$ & B4 & 256k & 0.01\\
A5 & 512k &     $0.001$ & B5 & 512k & 0.01\\
\hline
\end{tabular}\label{TableAB}
\end{table}

The $N$-body integrations were carried out using $\phi$GRAPE \citep{harfst}, a parallel, direct-summation $N$-body code that uses special-purpose hardware to compute pairwise forces between all particles:
\begin{equation}
\textbf{F}_{i} = -m_{i}\sum_{j=1,j \neq i}^{N}\frac{m_{j}(\textbf{r}_{i}-\textbf{r}_{j})}{(|\textbf{r}_{i}-\textbf{r}_{j}|^{2}+\epsilon^{2})^{3/2}}. \label{force}\\
\end{equation}
Here $m_{i}$ and $r_{i}$ are the mass and position of the \textit{i}th particle respectively and $\epsilon$ is a force softening parameter. 
We set $\epsilon = 10^{-5}$ in all the runs. $\phi$GRAPE integrates the equations of motion using a fourth-order Hermite integrator with individual block time steps 
similar to the well established Aarseth direct N-body codes\citep{maar92}. The time step of particle $i$ at time $t$ was computed from the standard formula:
\begin{equation}
\Delta t_{i} =\sqrt{\eta \frac{|\textbf{a}(t)||\textbf{a}^{2}(t)|+|\textbf{a}^{1}(t)|^{2}}{|\textbf{a}^{1}(t)||\textbf{a}^{3}(t)|+|\textbf{a}^{2}(t)|^{2}}} \label{tstep} \\
\end{equation}
where $\mathbf{a}$ is the acceleration of \textit{i}th particle and the superscript \textit{k} indicates the order of the time derivative. 
The time-step parameter $\eta$ was set to 0.01 for all our runs.
The choice of parameters was motivated by \citet{ber05} who found that integration accuracy is more sensitive to $\eta$ than to $\epsilon$. Those authors adopted $\eta = 0.01$ and $\epsilon = 1 \times 10^{-4}$ for their study of SMBH binary evolution. 
In order to resove the orbit of the massive binary in our simulations down to a semi-major axis $a\approx 10^{-4}$ we adopted the smaller softening length $\epsilon = 10^{-5}$. 
With these choices of $\eta$ and $\epsilon$, the relative energy error in our simulations was $\sim 10^{-8}$. 
For a more detailed discussion of these issues we refer the reader to section 2 of \citet{MMS2007} .

The $N$-body integrations were carried out on three, special-purpose computer clusters. 
{\tt Titan}, at the Astronomisches Rechen-Institut in Heidelberg, and {\tt gravitySimulator}, at the Rochester Institute of Technology are both 32-node clusters employing GRAPE accelerator boards \citep{harfst}.
Some calculations were also carried out on the GPU-enhanced cluster ``Kolob'' at the University of Heidelberg. For the latter runs we used the modified version of the $\phi$GRAPE code including the SAPPORO library \citep{gab09}.

\section{Isolated Models}\label{spherical-models}

Evolution of a binary SMBH in a gas-free galaxy can be divided into three phases \citep{BBR80}.
In first phase, the SMBHs are unbound to each other and move independently in the galaxy potential.  The evolution of the individual SMBH orbits during this phase is governed by dynamical friction against the background stars. The second phase begins when the two SMBHs are close enough together to form a bound pair. In this phase, the separation between the SMBHs decreases very rapidly due to the combined effects of dynamical friction and ejection of stars by the gravitational slingshot. 
(For a detailed discussion of this phase, we refer the reader to the careful analysis in \citet{MM2001}.) At the end of the second phase, the two SMBHs form a hard binary, defined as a binary that ejects passing stars with positive (unbound) energies. Phase three consists of slow evolution of the binary as depleted orbits are repopulated, via two-body scattering or some other process.
In a spherical galaxy with large $N$, the timescale for orbital re-population is essentially the two-body relaxation time which scales as $\sim N/\ln N$.
This is the origin of the ``final-parsec problem'' \citep{LR2005}.

\begin{figure}
\centerline{
  \resizebox{0.98\hsize}{!}{\includegraphics[angle=270]{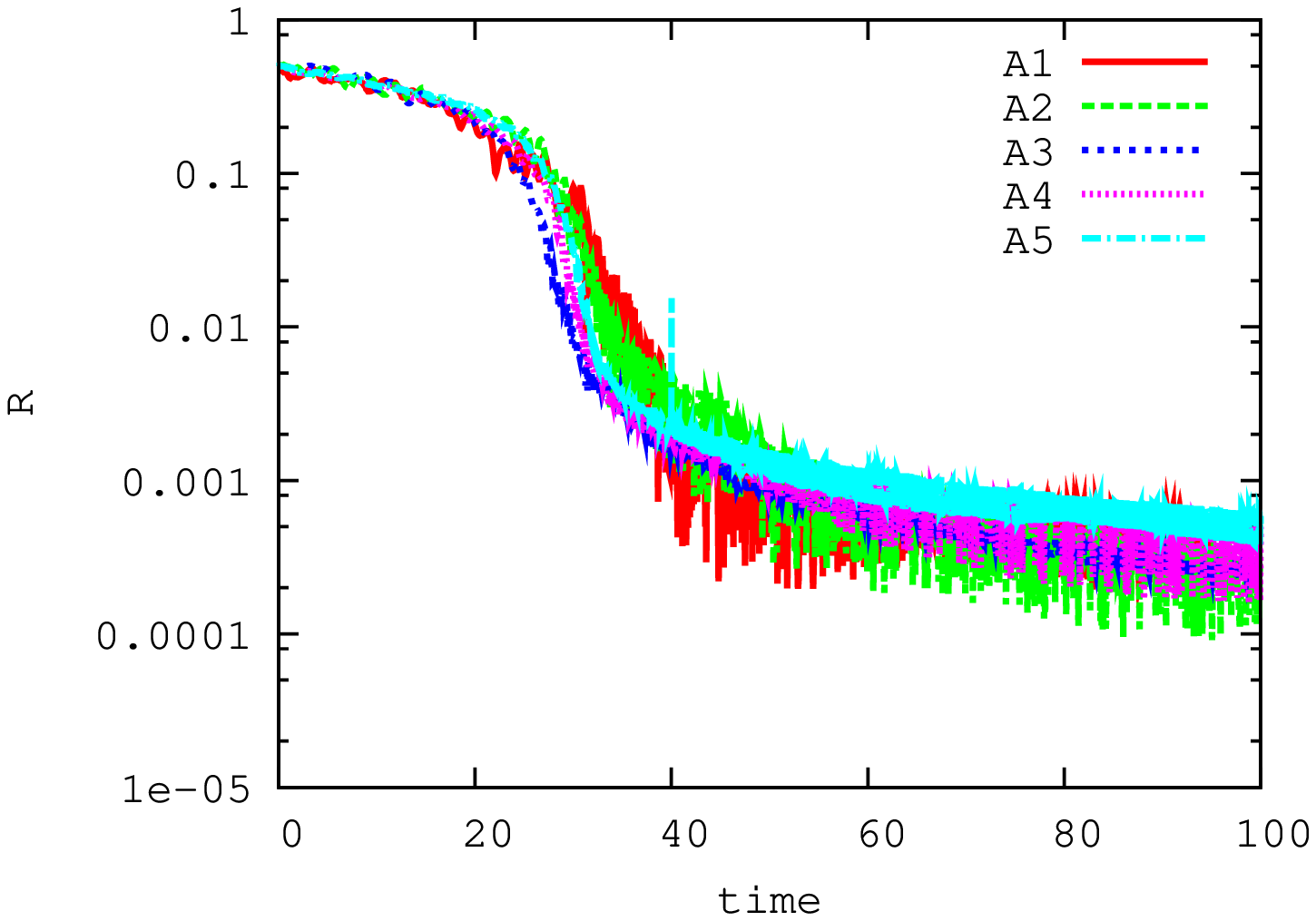}}
  }
\centerline{
  \resizebox{0.98\hsize}{!}{\includegraphics[angle=270]{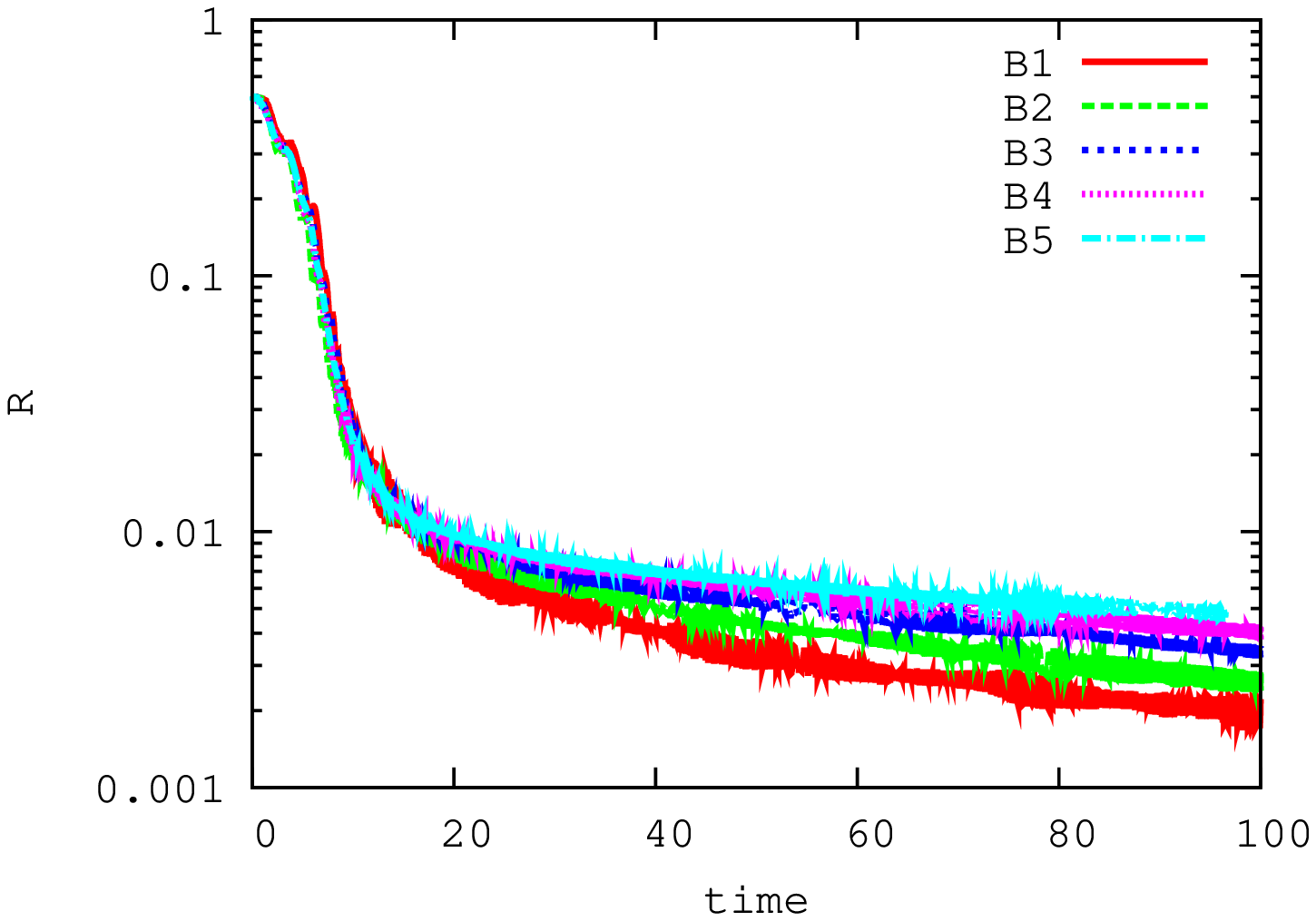}}
  }
\caption[]{
Evolution of the separation between the SMBHs, in ten $N$-body integrations of isolated spherical galaxies with different particle numbers according to Table \ref{TableAB}. Models A (top) have $M_\bullet=0.001$ and models B (bottom) have $M_\bullet=0.01$.
} \label{sepAB}
\end{figure}

Figure \ref{sepAB} plots the separation $R$ between the two SMBH particles versus time in the isolated galaxies, Models A and B. Here we can see the three phases of binary evolution described above. 

(1) In the first phase, the separation between the two SMBHs decreases due to dynamical friction. This phase ends when $R \approx r_{h}$, the gravitational influence radius. 
$r_{h}$ can be defined as the radius of a sphere around the binary SMBH that encloses a stellar mass equal to twice the binary mass.
For Models A, $M_{\bullet} = 0.001$ and $r_{h}\approx 0.07$. 
From Figure \ref{sepAB}, we can see that the first phase ends indeed when $R \approx 0.07$. Here we want to emphasize that this estimate does not take into account the changes in the galaxy that are induced by the presence of the second SMBH.

For Models B, the radius of influence is $\sim 0.25$. 
Judging from the bottom panel of Figure \ref{sepAB}, it appears that the first phase ends approximately when $R\approx 0.25$, although the transition is not as sharp as in model
A, presumably because the initial separation is only $\sim 2r_h$.

\begin{figure}
\centerline{
  \resizebox{0.98\hsize}{!}{\includegraphics[angle=270]{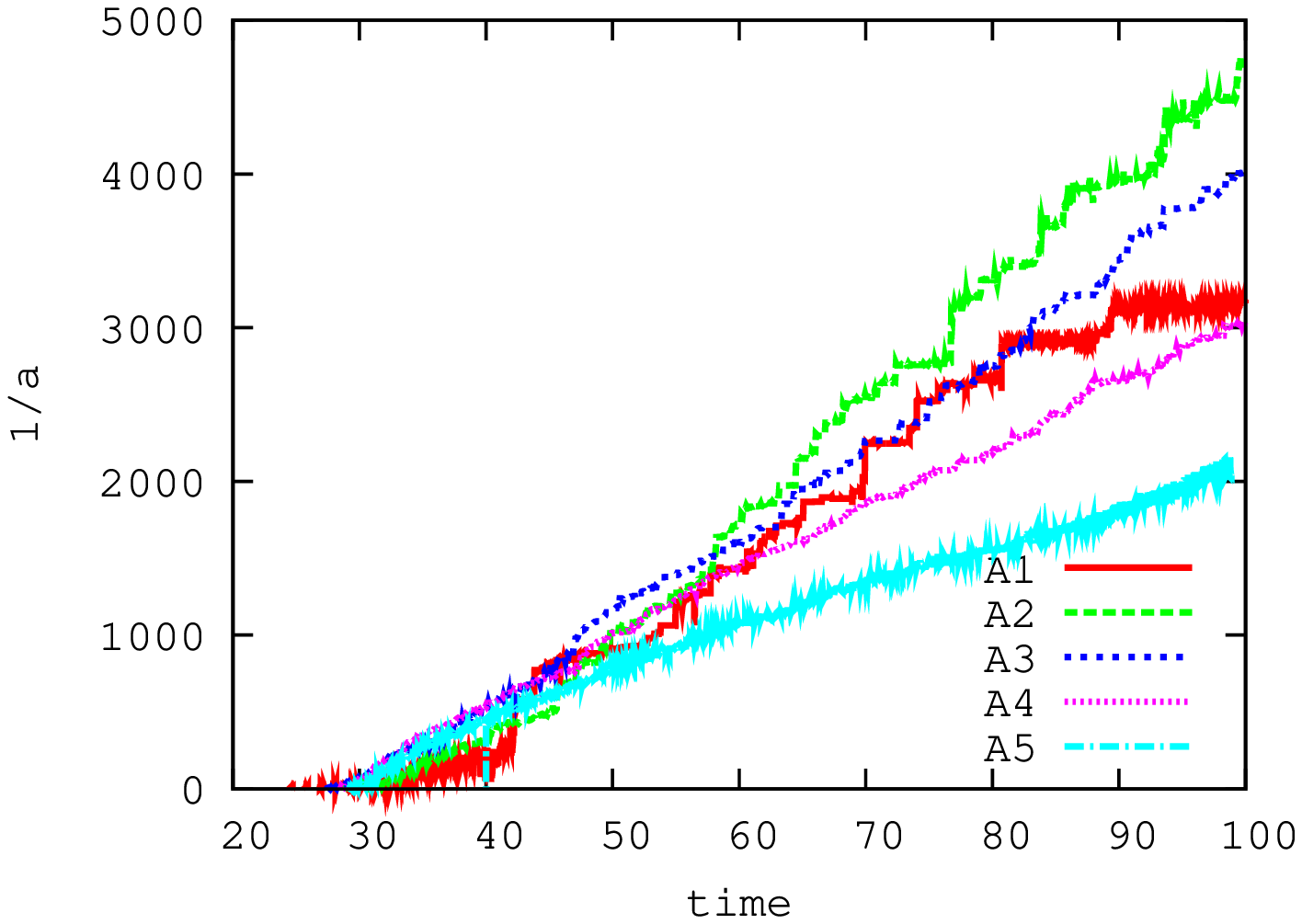}}
  }
\centerline{
  \resizebox{0.98\hsize}{!}{\includegraphics[angle=270]{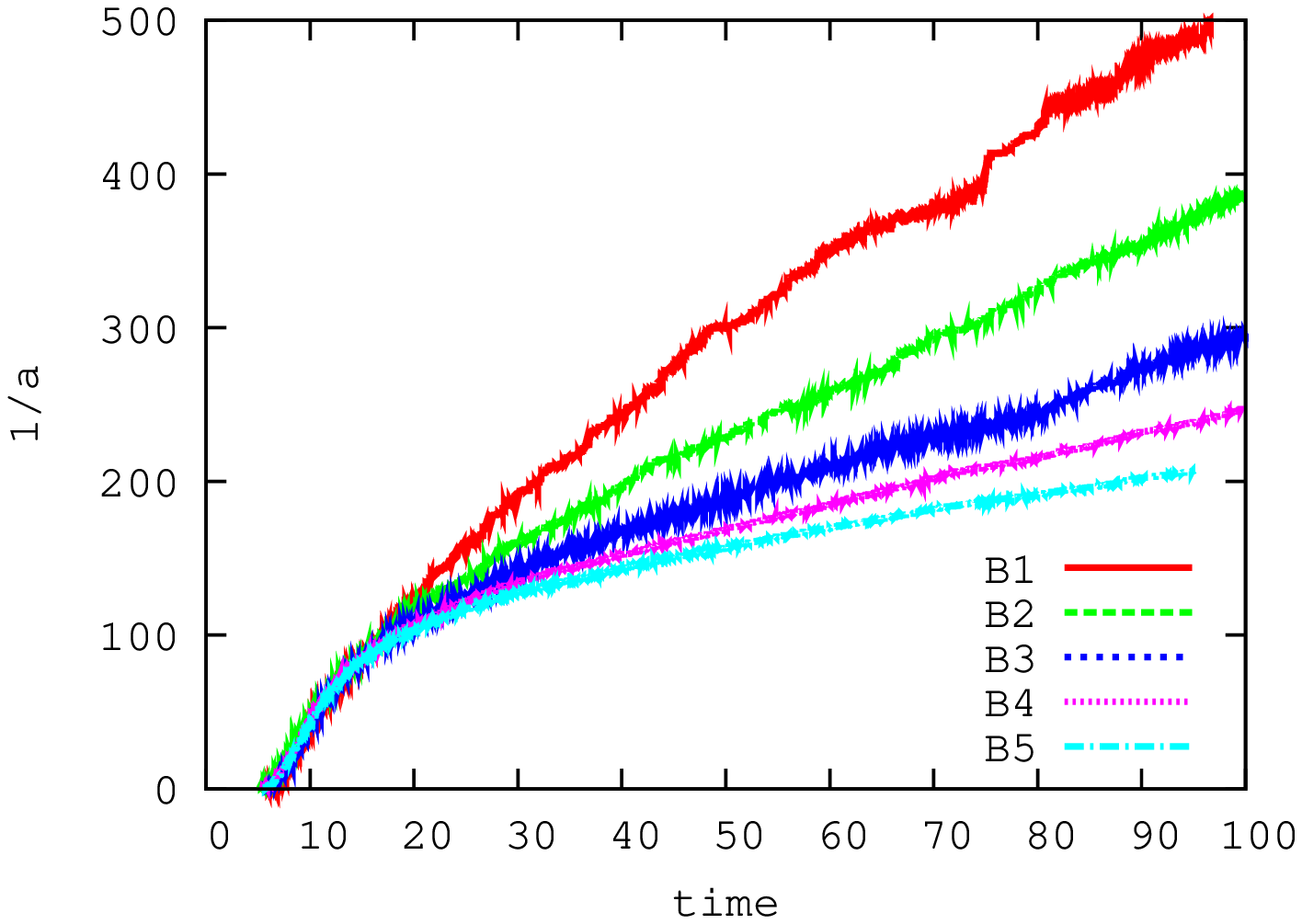}}
  }
\caption[]{
Evolution of the inverse semi-major axis of the massive binary in the isolated models.
} \label{semiAB}
\end{figure}

(2) In the second phase, the separation between the two SMBHs decreases very rapidly. 
Dynamical friction, and the ejection of stars by the gravitational slingshot, act together to efficiently extract angular momentum from the massive binary. 
These processes are not well defined in the regime where the massive binary is neither very hard nor very soft; equations (13) and (14) of \citet{MM2001} give approximate expressions for the rate at which energy is transferred to stars by the two mechanisms. 
For Models A this process continues until $t \approx 40$ while for Models B, this phase ends at $t \approx 20$.
The motion of two SMBHs in this phase is approximately Keplerian.  We define the semi-major axis $a$ and eccentricity $e$ via the standard relations:
\begin{equation}
\frac{1}{a} = \frac{2}{R} - \frac{v^{2}}{\mu},
\end{equation} 
\begin{equation}
e = \sqrt{1 + \frac{2h^{2}}{\mu ^{2}}\left[\frac{v^{2}}{2} - \frac{\mu}{R}\right]}.
\end{equation}
where  $\mu = G(M_{\bullet 1} + M_{\bullet2)}$, $v$ is the relative speed, and $h$ is the specific angular momentum of the relative motion.

(3) The rapid phase of binary hardening comes to an end when $a \approx a_{h}$, where $a_{h}$ is the semi-major axis of a ``hard binary'':
\begin{equation}
a_{h} = \frac{q}{(1+q)^{2}}\frac{r_{h}}{4}
\end{equation}
\citep[e.g.][] {MMS2007}.
Here $q = M_{\bullet 2}/M_{\bullet1})$ is the mass ratio of two SMBHs; for both Models A and B, $q = 1$.
For Models A, $a_{h} \approx 0.004$ with $a_{h}^{-1} \approx 250$ and for Models B, $a_{h} \approx 0.016$ with $a_{h}^{-1} \approx 65$.

From Figure \ref{semiAB}, we see that there is no clear dependence of the binary hardening rate on $N$ until the binary becomes hard, $ a < a_{h}$. 
For Models A this happens around $t = 25$ and for Models B around $t = 30$.  Beyond these times the binary hardening rate exhibits a clear $N$ dependence in the sense that hardening is slower for larger $N$. Similar $N$-dependence has been seen in other studies of binary evolution in spherical galaxies \citep[e.g.][]{MF2004,ber06,MMS2007}.
For real spherical galaxies, with much larger $N$, the binary would stop evolving beyond this point.  

\begin{figure}
\centerline{
  \resizebox{0.98\hsize}{!}{\includegraphics[angle=270]{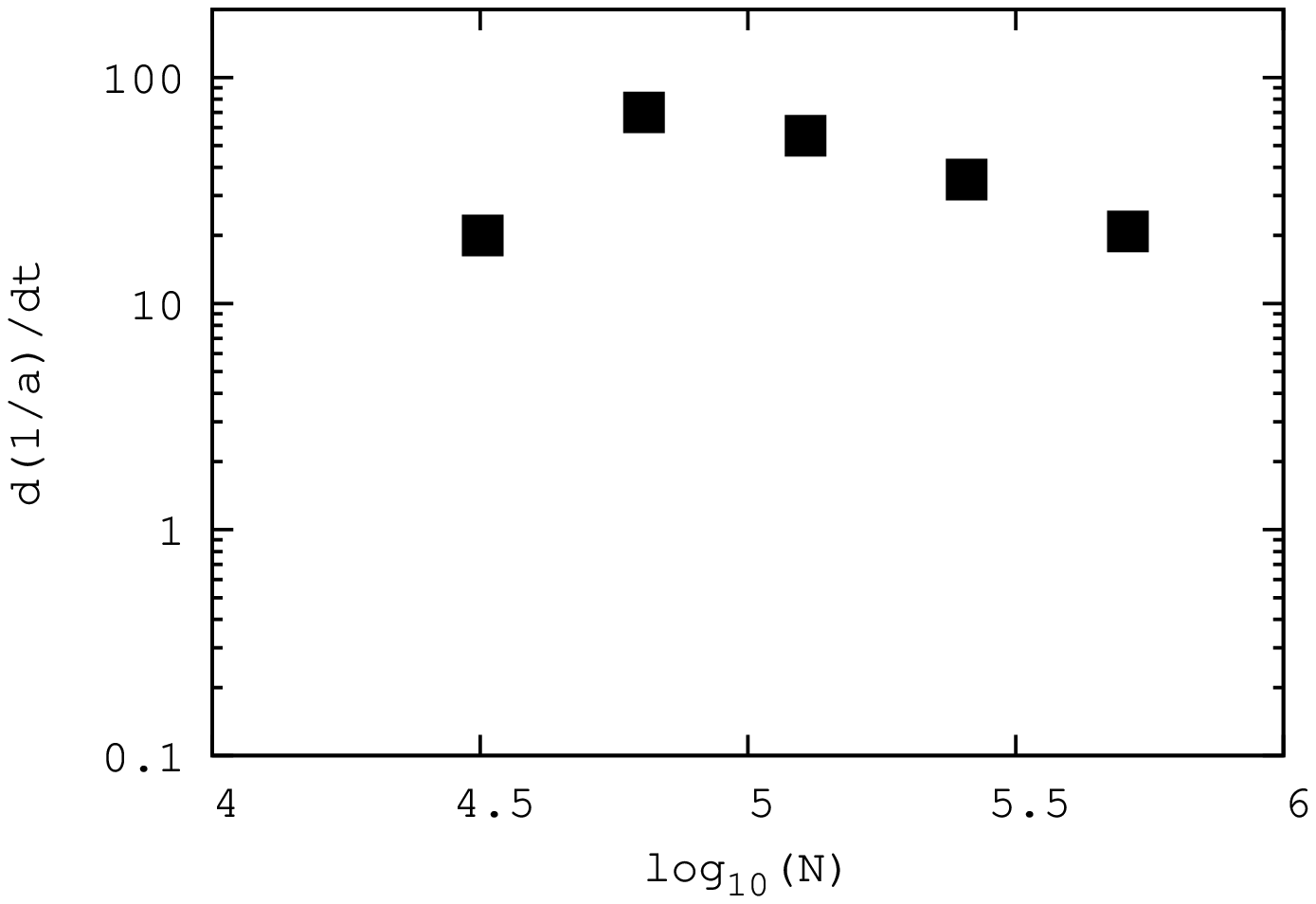}}
  }
\centerline{
  \resizebox{0.98\hsize}{!}{\includegraphics[angle=270]{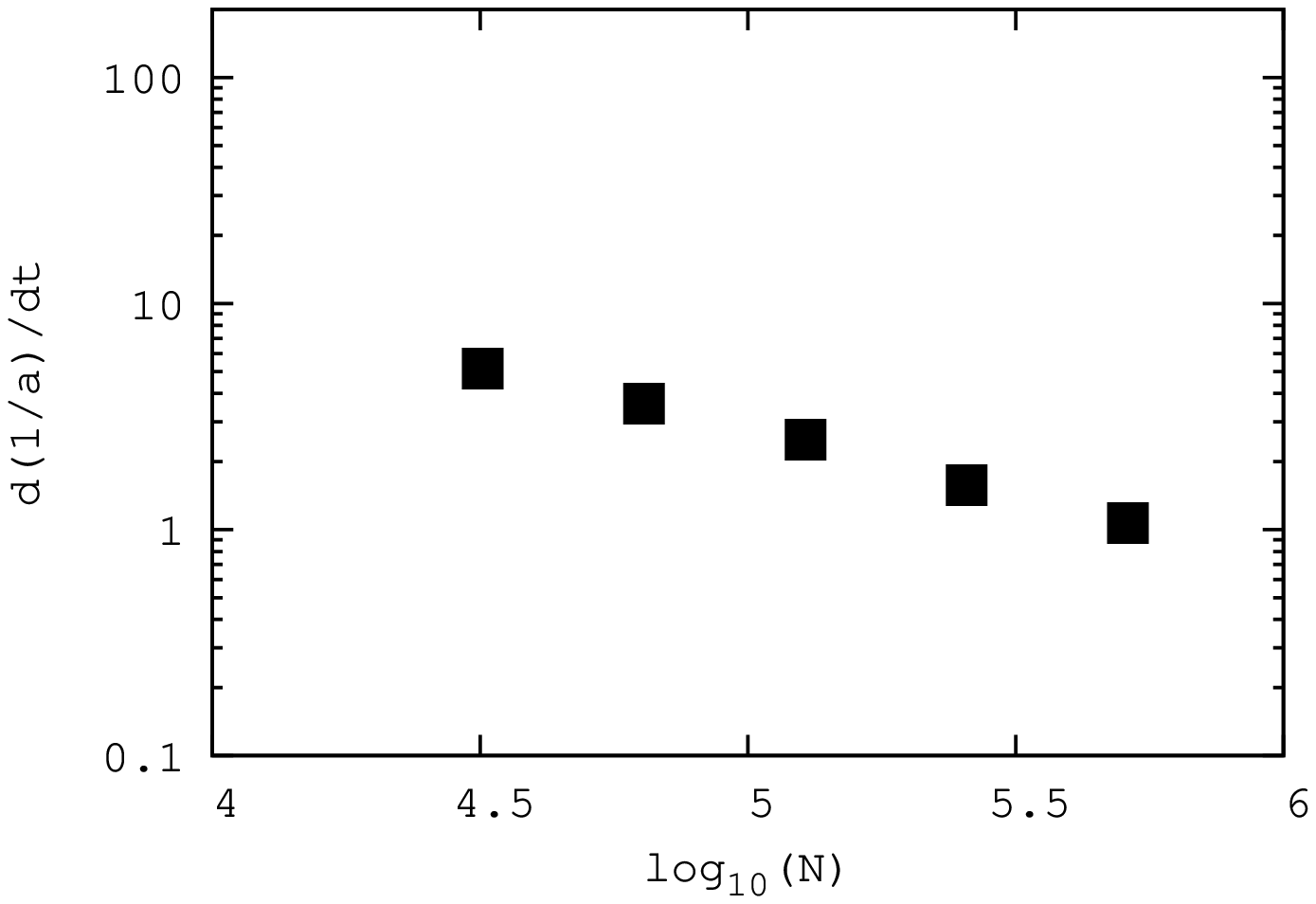}}
  }
\caption[]{
$N$-dependence of the binary hardening rate in the isolated galaxy models.
{\it Top:} Models A; {\it bottom:} Models B.
} \label{rateAB}
\end{figure}

We estimated the hardening rate $s$ in the $N$-body integrations, where
\begin{equation}
s \equiv \frac{d}{dt}\left(\frac{1}{a}\right), \label{speed}
\end{equation}
by fitting straight lines to $a^{-1}(t)$ in an interval $\Delta t = 50$ from $t = 50$ to $t = 100$. 
Figure \ref{rateAB} shows the $N$-dependence of $s$.
We see that for $N > 50$k, the hardening rate is a decreasing function of $N$;
scaling as $\propto N^{-0.6}$; this is consistent with the scalings found by other authors in the same $N$-range, e.g. \citet{ber05,MMS2007}.
Comparison of Figures~\ref{rateAB}a and b also shows lower hardening rates for Model B compared with Model A, by roughly an order of magnitude.
For large $N$, the two-body relaxation time is long compared with orbital periods of stars near the massive binary, and the loss cone around the binary is nearly empty. 
In this regime, and in the asymptotic (large-$N$) limit, the rate of binary evolution is 
predicted to scale as  $\sim \left[(M_{\bullet1}+M_{\bullet2})\right]^{-1}$
(e.g. equation~(32) of \citet{MMS2007}).
Given that $a$ is roughly a factor ten smaller, at a given
time, in Models A as compared with Models B,
there is an additional delay in the early phase of the hard binary evolution. However this delay is negligible for the
decay time $t_0$ until which the semimajor axis reaches $a_0$, when energy loss by GWs start to dominate. In section \ref{sec-disc} a more detailed discussion of the total time to merge the binary SMBH is given.


Figure \ref{den} shows the density profile for model A5 in the final phase. 
The central logarithmic slope of the density profile has dropped from $-1$ to $\sim -0.6$.

\section{Galaxy Mergers} \label{sec-merger}

We studied the evolution of binary SMBHs in merging galaxies by creating two identical galaxy models, each with a central SMBH, and placing the two galaxies on bound relative orbits. 
Table \ref{TableCD} summarizes the parameters of the galaxy merger models. 
The initial separation of the two galaxy centers was $20r_{0}$.  The initial relative velocity of the two galaxies was chosen such that the SMBH separation at first pericenter passage was $\sim r_{0}$; in other words, the initial orbit of the binary galaxy was substantially eccentric.  This was done in order to reduce the computational time required to bring the two SMBHs close together; in fact, a galaxy merger from such nearly ``head-on'' initial conditions is unlikely.

\begin{table}
\caption{Parameters of the galaxy merger simulations} 
\begin{tabular}{c c c }
\hline
Run & 2 $\times$ $N$ & $M_\bullet$ \\
\hline
C1 & 64k &	   $0.001$ \\
C2 & 128k &        $0.001$ \\
C3 & 256k &        $0.001$ \\
C4 & 512k &        $0.001$ \\
D1 & 64k  &	   $0.01$ \\
D2 & 128k &        $0.01$ \\
D3 & 256k &        $0.01$ \\
D4 & 512k &        $0.01$ \\
\hline
\end{tabular}\label{TableCD}
\end{table}

\begin{figure}
\centerline{
  \resizebox{0.98\hsize}{!}{\includegraphics[angle=270]{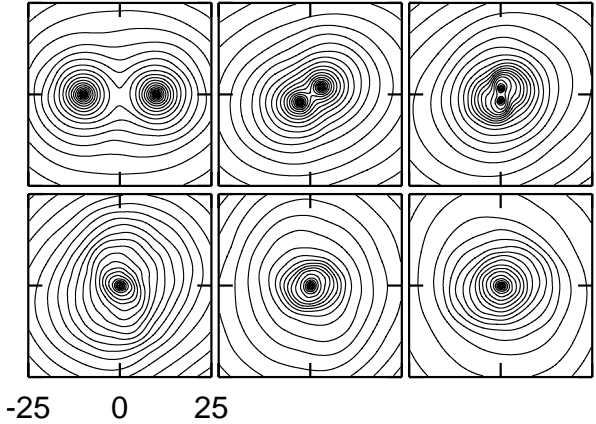}}
  }
\caption[]{
Density contours projected onto the initial orbital plane for Model C3. 
First row: $t=( 0, 70, 90)$. Second row: $t=(110, 150, 200)$. 
} \label{contour}
\end{figure}

\begin{figure*}
\centerline{
  \resizebox{0.48\hsize}{!}{\includegraphics[angle=270]{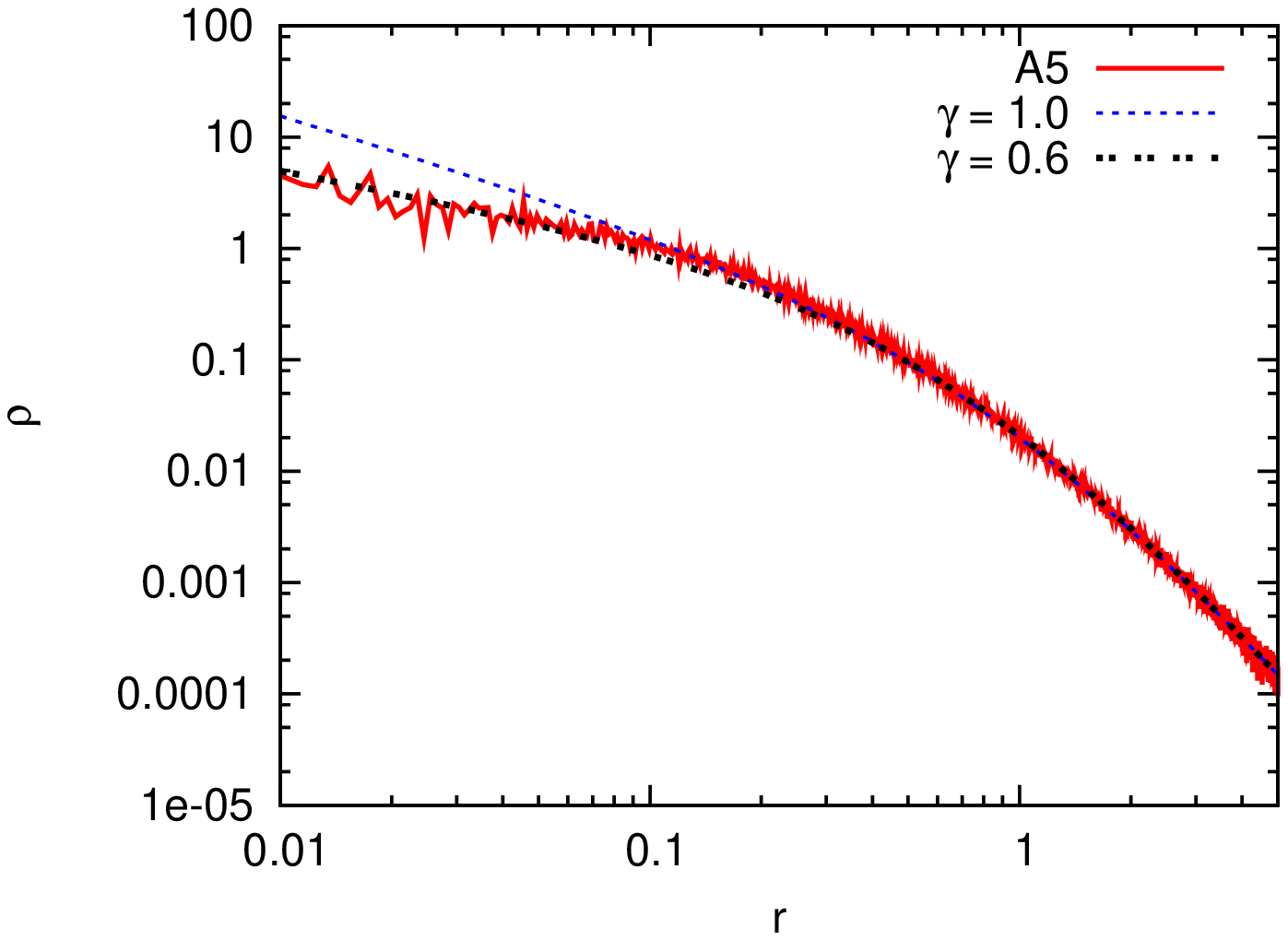}}
  \resizebox{0.48\hsize}{!}{\includegraphics[angle=270]{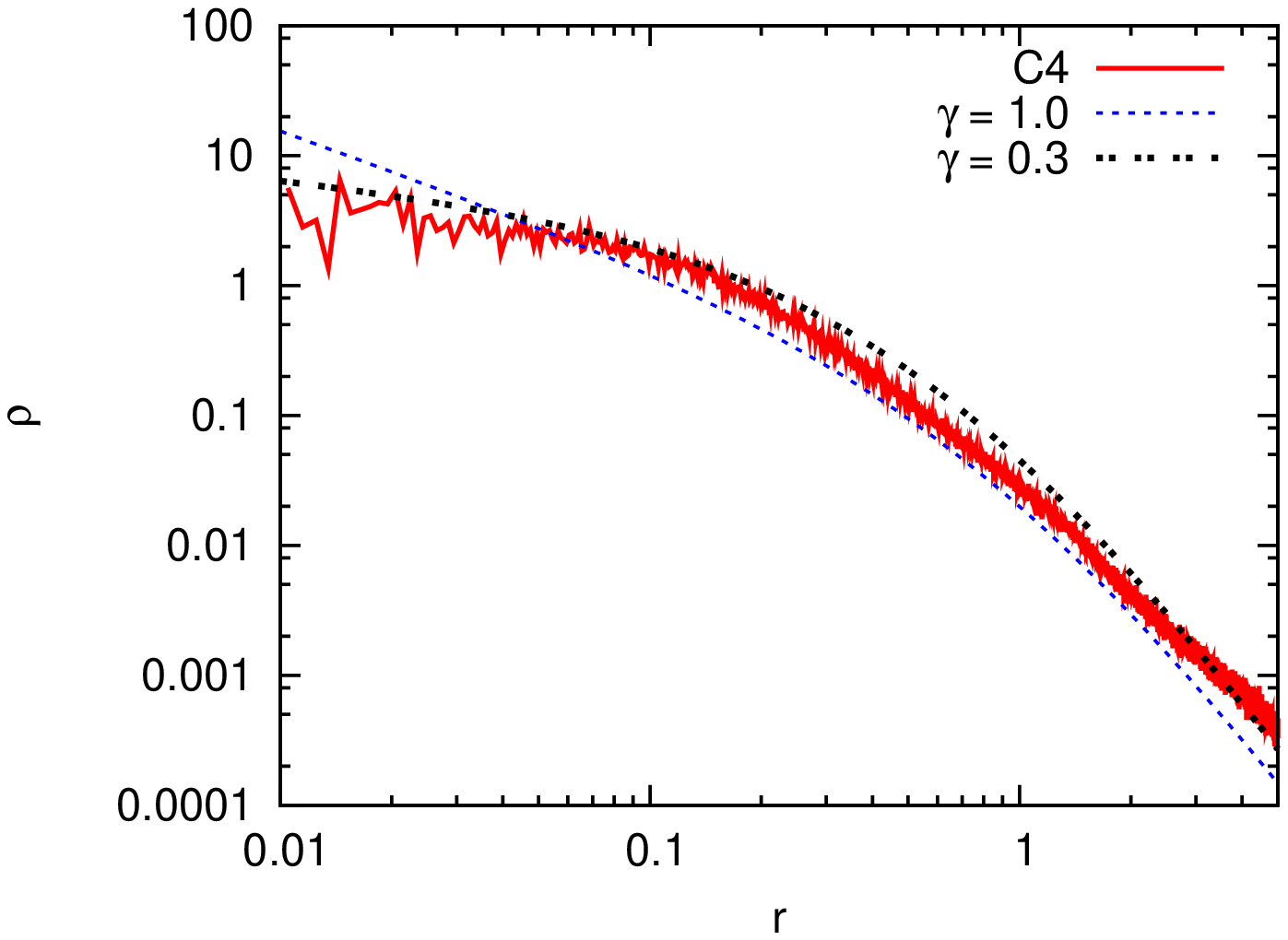}}
  }
\caption[]{
{\it Left:} Spatial density profile for model A5 averaged over five different time steps ($t = 60, 68, 76,84, 92$). The figure also shows the initial density profile ($\gamma = 1$) and a fit to the data using Eq.~\ref{denr} with $\gamma = 0.6$ and $r_{0} = 0.8$. The central density drops as the binary ejects mass from the core. 
{\it Right:}
Spatial density profile for model C4 averaged over five different time steps ($t = 140, 160, 180, 200, 220$). The figure also shows the initial density profile ($\gamma = 1$) and a fit to the data using equation \ref{denr} with $\gamma = 0.3$ and $r_{0} = 0.6$. 
} \label{den}
\end{figure*}

Figure \ref{contour} illustrates the merging of two galaxies. 
In the initial phase (top panels) the two SMBH particles remain strongly associated 
with their respective density cusps. 
The two cusps merge into one at $t \sim 100$, following which the cusp is gradually
destroyed by the gravitational slingshot ejection of stars (Fig.~\ref{den}).

\begin{figure}
\centerline{
  \resizebox{0.98\hsize}{!}{\includegraphics[angle=270]{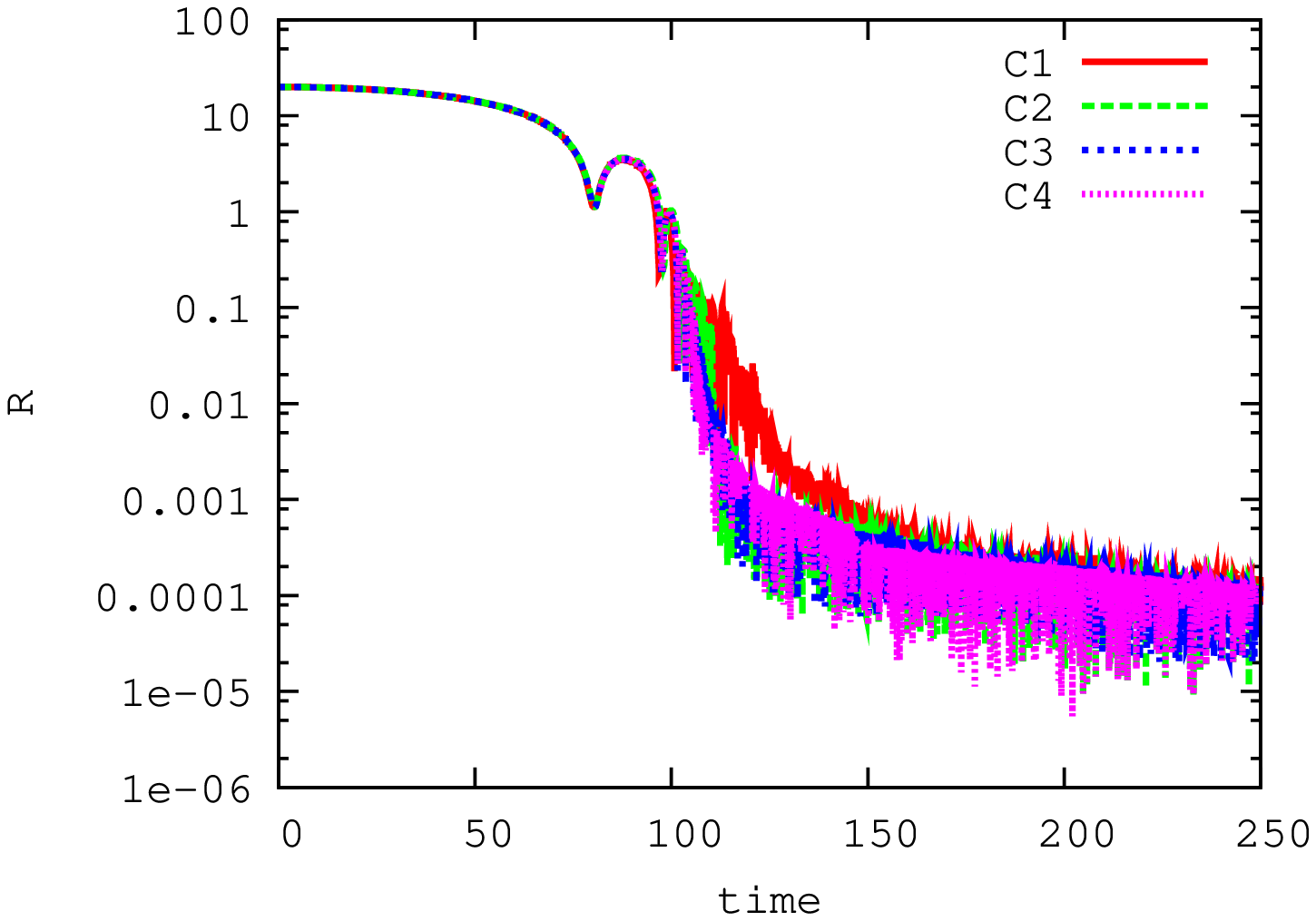}}
  }
\centerline{
  \resizebox{0.98\hsize}{!}{\includegraphics[angle=270]{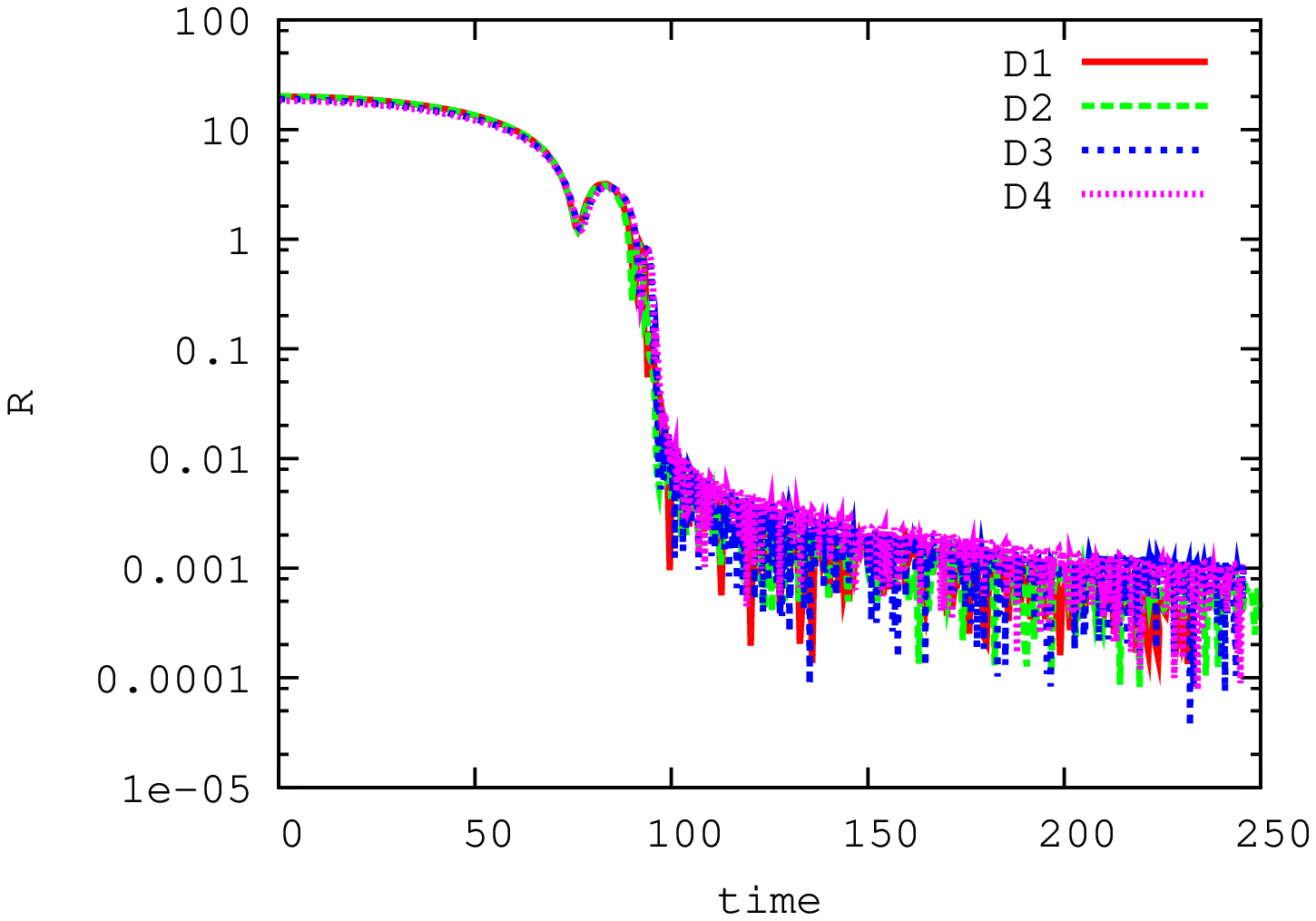}}
  }
\caption[]{
Evolution of the separation between the two SMBHs in the galaxy merger simulations.
} \label{sepCD}
\end{figure}

Figure \ref{sepCD} shows the separation between the two SMBHs in the merger models. The first pericenter passage occurs around $t \approx 80$.   As the galaxies merge, the two SMBHs remain centrally located in their respective density cusps, until coming close enough together that a binary SMBH forms, at $t \approx 100$, or slightly earlier in the case of Model D (phase one).
The separation between the two SMBHs then decreases very rapidly due to the combined effects of dynamical friction and slingshot ejection of stars (phase two). Once the hard binary is formed, the hardening rate of the binary decreases (phase three). This behavior is qualitatively similar to what was seen in the isolated galaxy models.

\begin{figure}
\centerline{
  \resizebox{0.98\hsize}{!}{\includegraphics[angle=270]{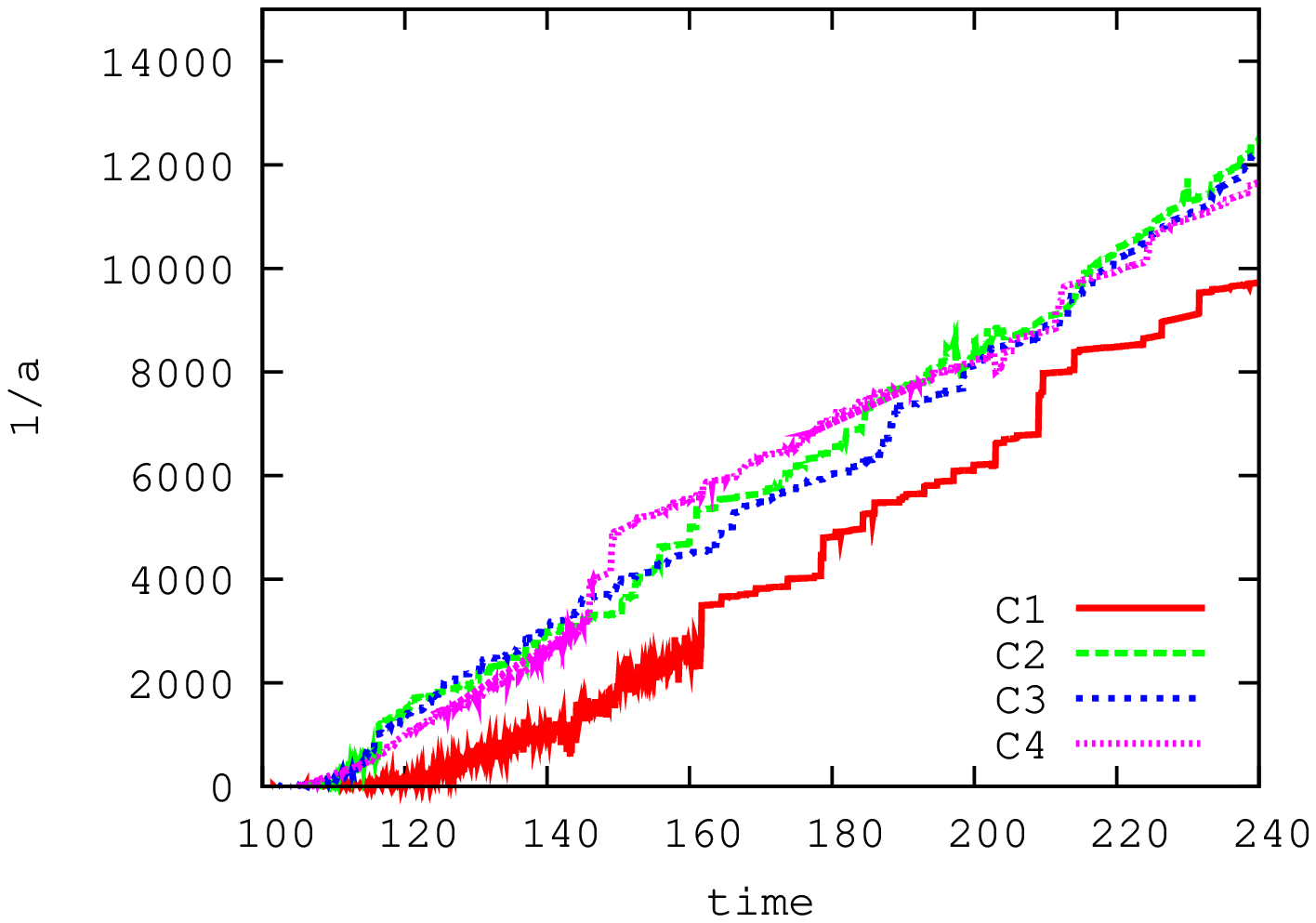}}
  }
\centerline{
  \resizebox{0.98\hsize}{!}{\includegraphics[angle=270]{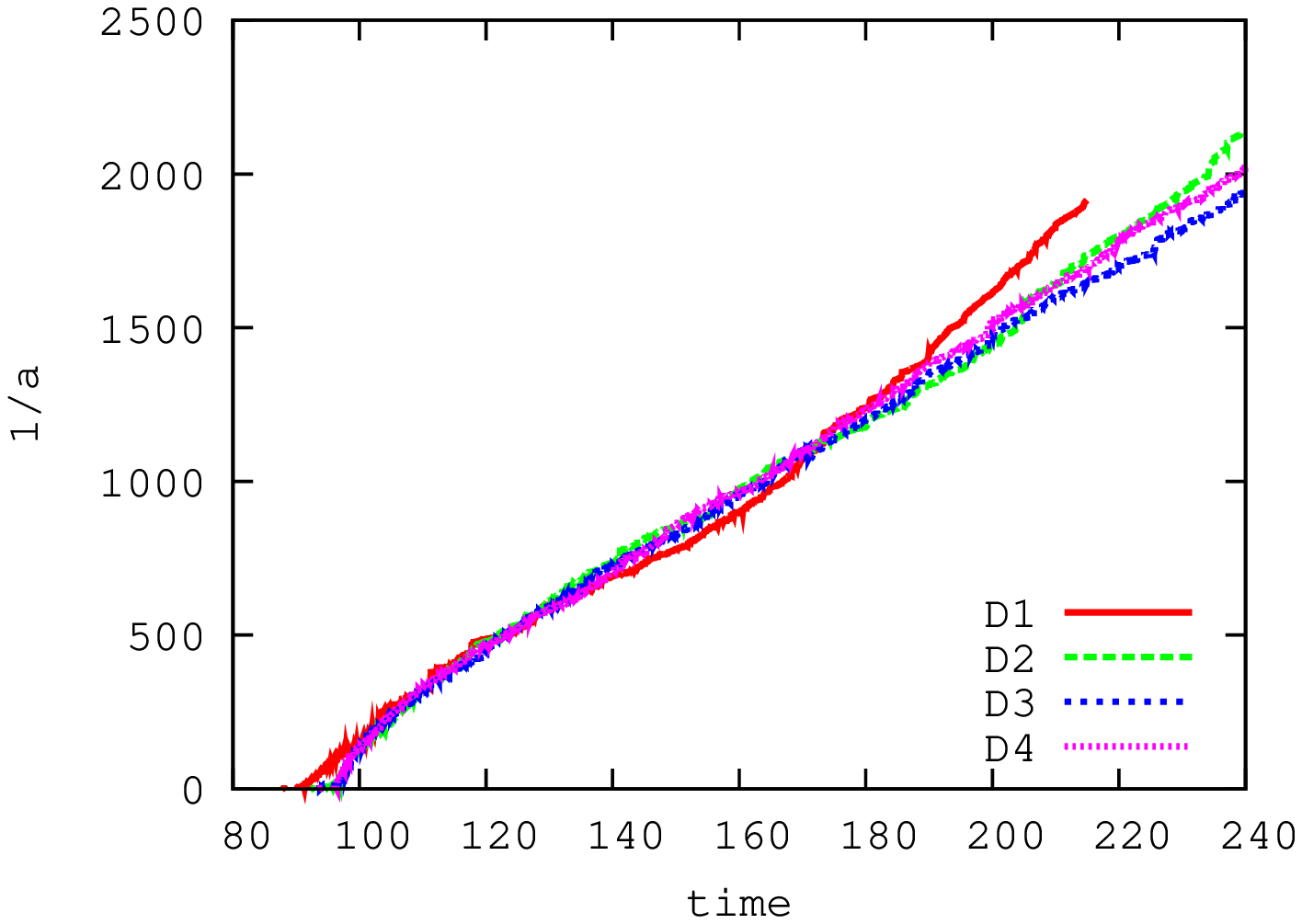}}
  }
\caption[]{
Evolution of the inverse semi-major axis of the binary SMBH during the galaxy merger simulations.
} \label{majCD}
\end{figure}

\begin{figure}
\centerline{
  \resizebox{0.98\hsize}{!}{\includegraphics[angle=270]{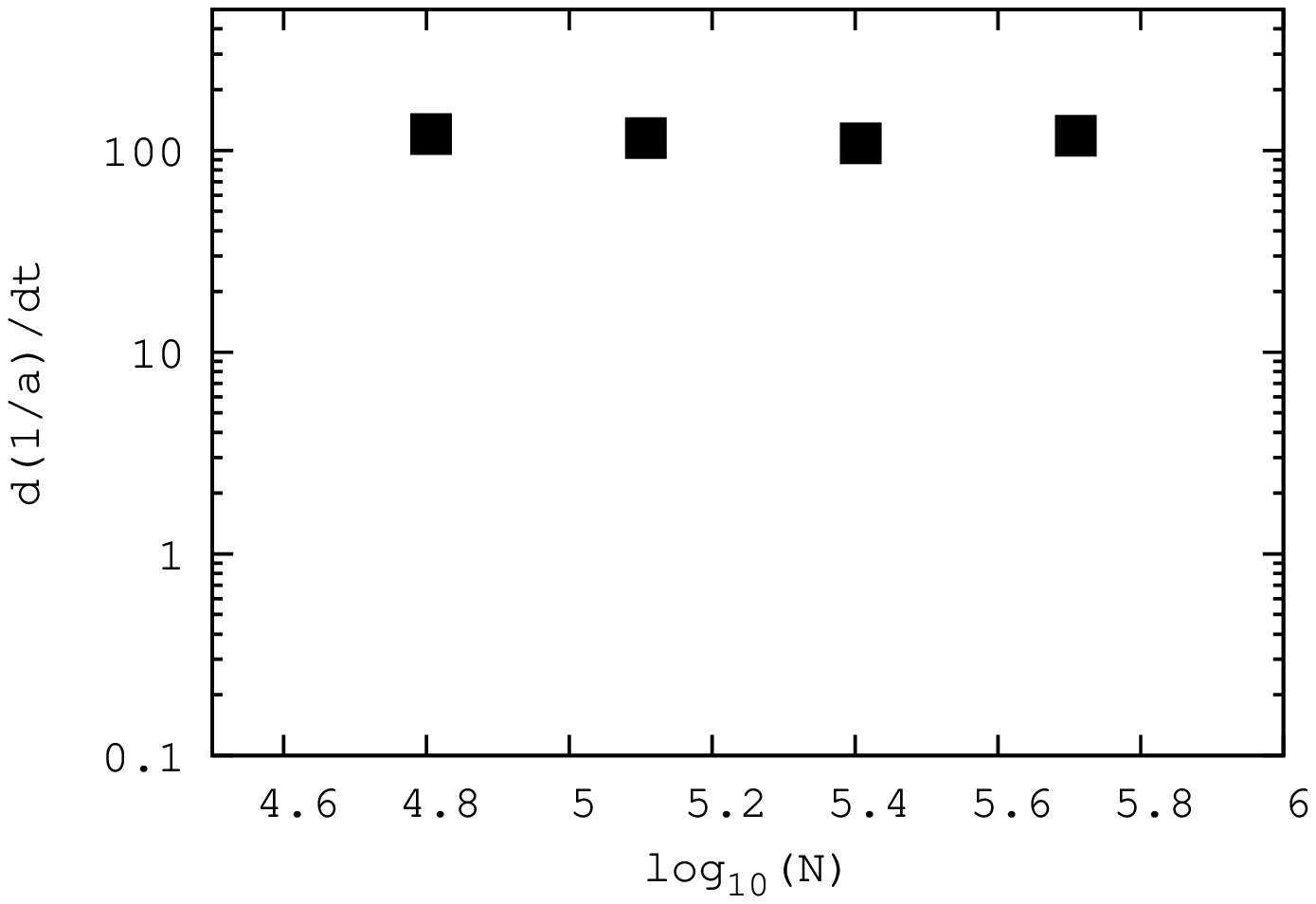}}
  }
\centerline{
  \resizebox{0.98\hsize}{!}{\includegraphics[angle=270]{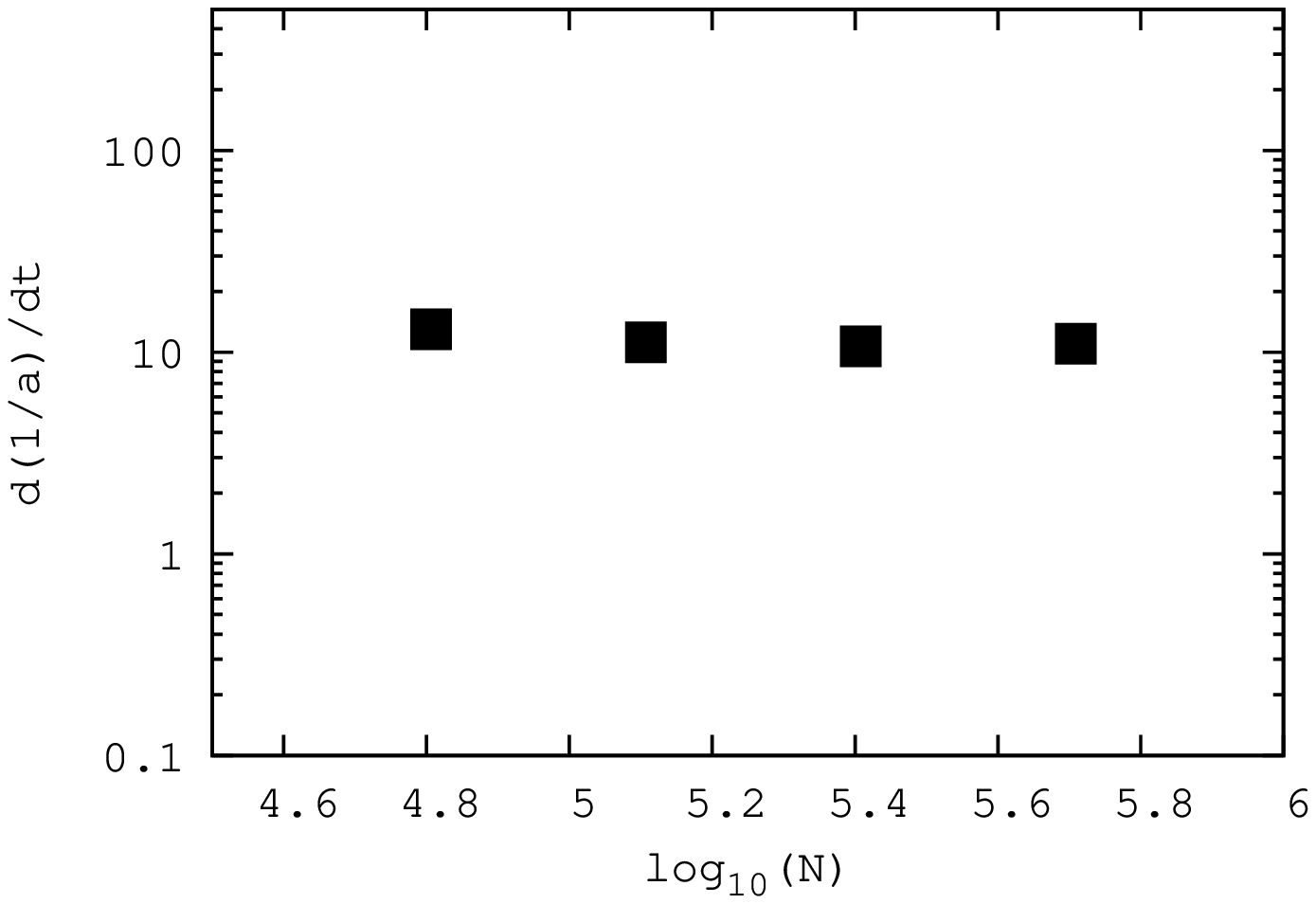}}
  }
\caption[]{
Binary hardening rates in the galaxy merger simulations (top panel: Model C, bottom panel: Model D).
} \label{rateCD}
\end{figure}

However there is one important difference between the binary evolution in the isolated and merging galaxies.
Figure~\ref{majCD} shows the evolution of the binary semi-major axis, and 
Figure~\ref{rateCD} shows the binary hardening rates in the galaxy merger simulations; the latter were computed in the same way as in the isolated galaxy models, by linear fit of $1/a$ in the range $150 \le t \le 200$. 
In the galaxy merger models, there is essentially no dependence of the binary hardening rate on particle number.
Furthermore, in Models C and D, the hardening rates are much higher -- more than five times higher than in the Models A and B, respectively, when $N=512$k.

A likely explanation for this difference is that the non-spherical shapes of the
merger remnants provide an additional source of torque on the stellar orbits,
allowing them to interact with the central binary on a timescale shorter than
the two-body relaxation time.
It has been argued \citep{MP2004,ber06} that allocating even a small fraction
of the stellar orbits to a ``centrophilic'' family could lead to binary
hardening rates that are much larger than those that result from two-body
relaxation alone.

\begin{figure}
\centerline{
  \resizebox{0.98\hsize}{!}{\includegraphics[angle=270]{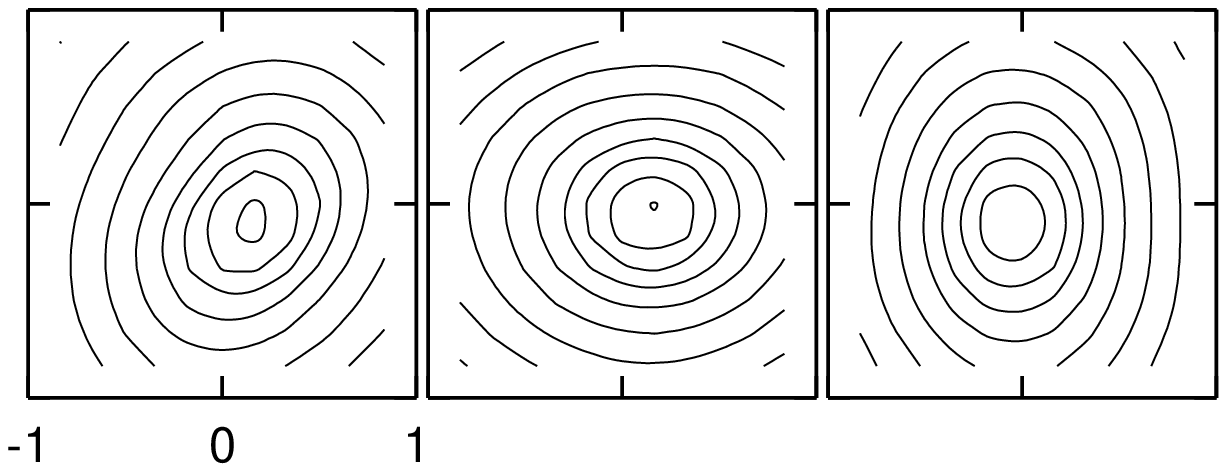}}
  }
\caption[]{
Density contours for merger Model C3 at t = 150, projected on xy plane(left), xz plane(center), yz plane(right). 
} \label{contour1}
\end{figure}

\begin{figure}
\centerline{
  \resizebox{0.98\hsize}{!}{\includegraphics[angle=270]{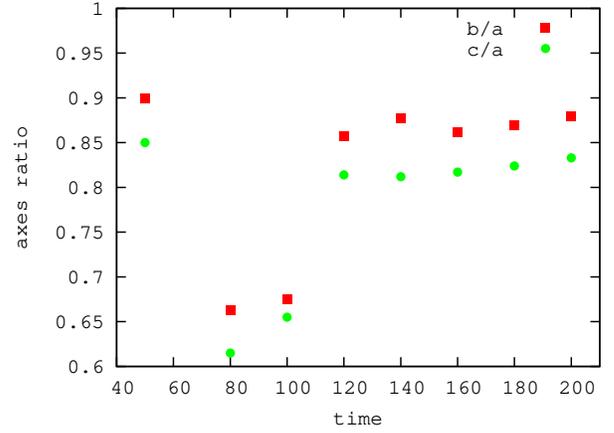}}
  }
\caption[]{
Ratio of intermediate to major (b/a) and minor to major (c/a) axes for Run C3. 
} \label{axisC}
\end{figure}

We evaluated the shapes of the merged galaxies in a number of ways.
The central density contours for Model C3 are shown in Figure~\ref{contour1}. 
Departures from axisymmetry are evident. 
Figure~\ref{axisC} shows the principle axis ratios of the merger remnant in Model C3 as a function of time; these were defined as the axis ratios of a homogeneous
ellipsoid with the same inertia tensor.
The departures from spherical symmetry are modest, but definite, and they
appear to be nearly independent of time toward the end of the simulation.

\begin{figure}
\centerline{
  \resizebox{0.98\hsize}{!}{\includegraphics[angle=270]{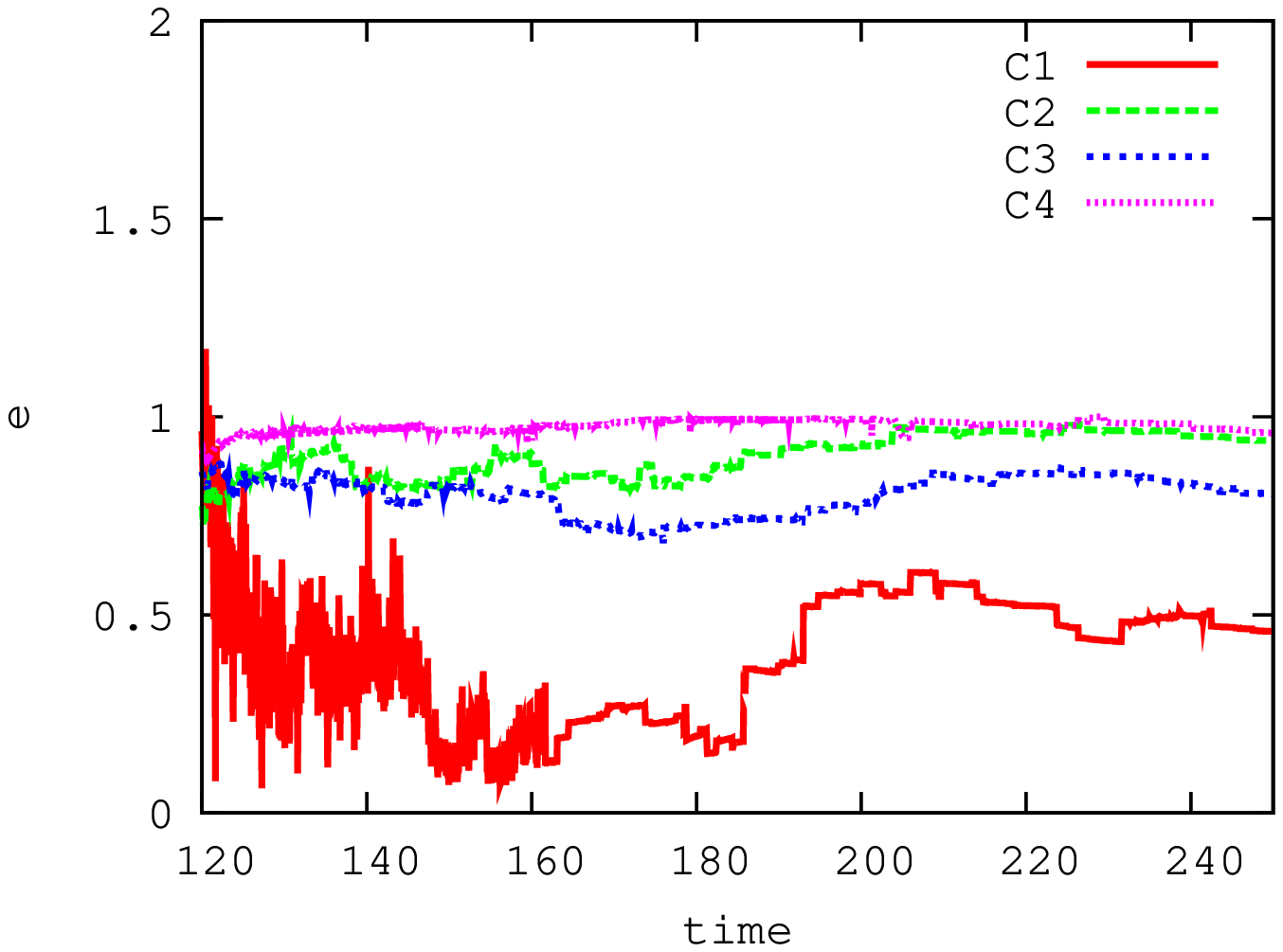}}
  }
\centerline{
  \resizebox{0.98\hsize}{!}{\includegraphics[angle=270]{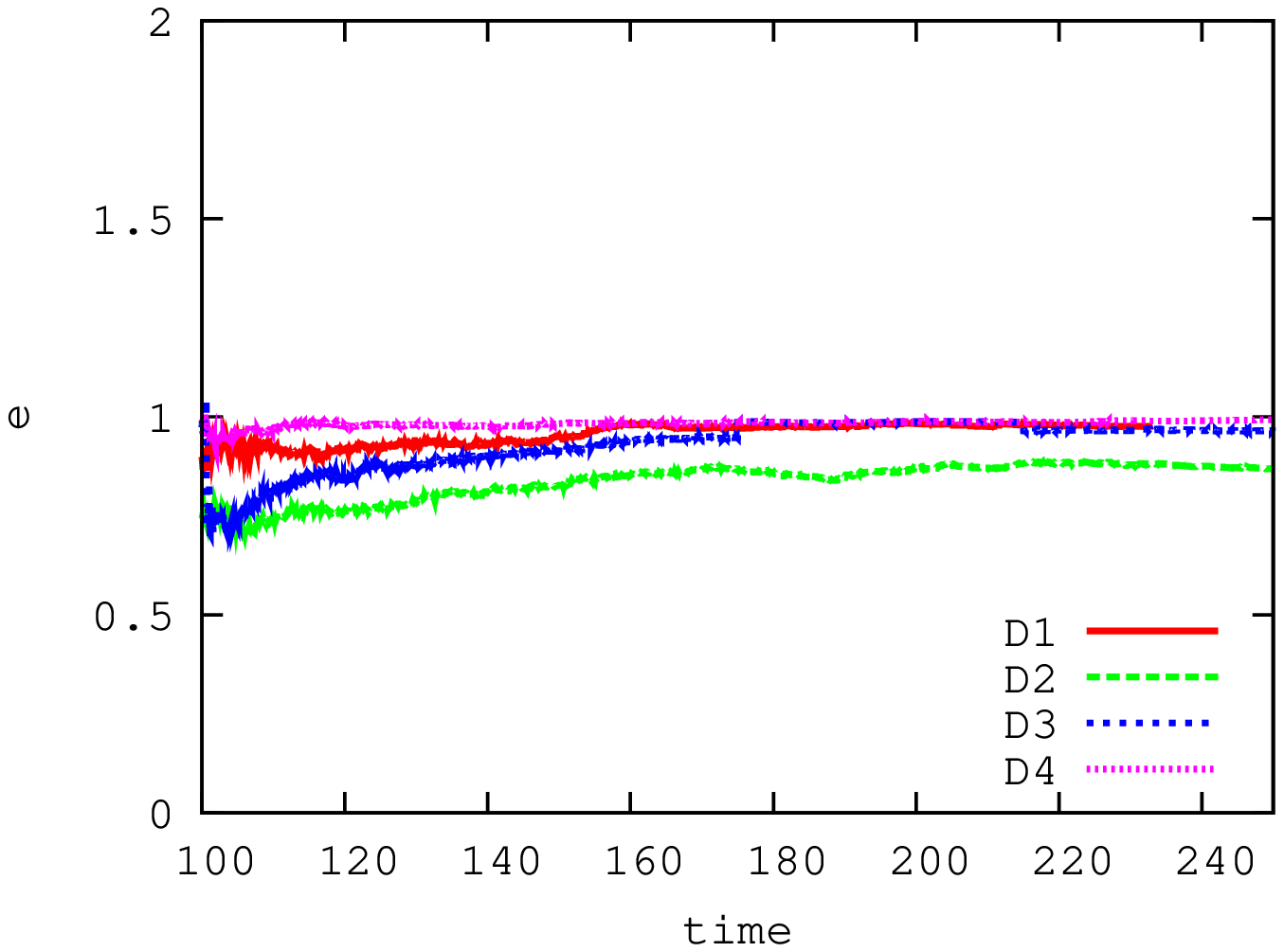}}
  }
\caption[]{
Evolution of binary eccentricity for models C (top) and models D (bottom).
} \label{eccCD}
\end{figure}

The evolution of binary eccentricity in the merger models is presented in Figure~\ref{eccCD}. 
We notice high eccentricities for both models as soon as the SMBHs become bound, approaching unity for models D. 
The high eccentricities are due in part to the high eccentricity ($e=0.95$) of the galaxies' relative orbit prior to the merger, and, in some runs, to post-merger evolution. 

\section{Time Scales for Coalescence} \label{sec-disc}

SMBH binaries are a potentially important source of gravitational wave (GW) emission. Gravitational waves extract energy and angular momentum
from the binary, hence changing its orbital elements.
\citet{P1964} gives approximate, orbit-averaged expressions for the rates of change of a binary's semimajor axis and eccentricity due to GW emission:
\begin{subequations}
\begin{eqnarray}
\left(\frac{da}{dt}\right)_\mathrm{GW}  &=& -\frac{64}{5}\frac{G^{3}M_{\bullet1}M_{\bullet2}(M_{\bullet1}+M_{\bullet2})}{a^{3}c^{5}(1-e^{2})^{7/2}}\times \nonumber \\
&&\left( 1+\frac{73}{24}e^{2}+\frac{37}{96}e^{4}\right),  \label{dadt}\\
\left(\frac{de}{dt}\right)_\mathrm{GW}  &=& -\frac{304}{15}e\frac{G^{3}M_{\bullet1}M_{\bullet2}(M_{\bullet1}+M_{\bullet2})}{a^{4}c^{5}(1-e^{2})^{5/2}}
\times\nonumber\\
&&\left( 1+\frac{121}{304}e^{2}\right) .  \label{dedt}
\end{eqnarray}
\end{subequations}
 We define $t_\mathrm{GW}(a_0,e_0)$ as the time required,
according to the coupled equations (\ref{dadt},~\ref{dedt}),
for the binary semi-major axis to shrink to zero under the influence of GW
emission, starting from the initial values $(a_0, e_0)$.
This time is a strong function of $e_0$ for $e_0\approx 1$, as it is in some
of our simulations.

The full time to coalescence, $t_\mathrm{coal}$, includes also the 
time from the start of the simulation until the GW regime is entered.
We estimated $t_\mathrm{coal}$ as follows.
In the $N$-body merger models, the post-merger
hardening rate of the binary, $s$, is essentially independent of $a$ and $N$
(Figures~\ref{majCD}, \ref{rateCD}).
The rate of change of the binary semi-major axis in this regime is 
approximately
\begin{equation}\label{eq:dadtnb}
\left(\frac{da}{dt}\right)_\mathrm{NB} \approx -sa^2, \ \ s\approx \mathrm{const.}
\end{equation}
Given values for $M_\mathrm{gal}$ and $r_0$, this dimensionless
rate can be converted into physical units.
At some time in the simulation, $a$ will have fallen to such a small value that
$(da/dt)_\mathrm{NB}$ as given by equation~(\ref{eq:dadtnb}) will equal $(da/dt)_\mathrm{GW}$, where the latter quantity is computed from
{\it both} $a$ and $e$.
We define $t_0$ as the time at which these two rates are equal.
The full time to coalescence, $t_\mathrm{coal}$, is the time from the
start of the simulation until $t_0$, plus $t_\mathrm{GW}(a_0,e_0)$,
where $a_0\equiv a(t=t_0), e_0\equiv e(t=t_0)$.

In most cases, $t_0$ exceeded the final time $t_\mathrm{final}$ of the simulation.
In these cases, the time $\Delta t$ between $t_\mathrm{final}$ and $t_0$ was assumed to be
 \begin{equation}
\Delta t = s(t_\mathrm{final})^{-1}\left(\frac{1}{a_0} - \frac{1}{a_\mathrm{final}}\right).
 \label{tcoal}
\end{equation}
In such cases, we also assumed that $e(t)$ was constant, $e(t)=e(t_\mathrm{final})$,
for $t>t_\mathrm{final}$  when computing $t_\mathrm{GW}$.

Computation of the GW evolution rates required adoption of particular values
for the physical units of length and time, hence of $M_\mathrm{gal}$ and $r_0$.
Table 3 gives results for two different assumed values of the galaxy mass:
$M_\mathrm{gal}=(10^9,10^{11})M_\odot$.

Because of the strong dependence of $t_\mathrm{GW}$ on eccentricity,
Table 4 gives $a_\mathrm{GW}$ and $\Delta t$, for run C3 only,
under two different assumptions about $e_0$: $e_0=0.5$ and $e_0=0$.
The total decay time increases by a factor of 2--3 if the orbit is assumed circular
at the start of the GW phase.

For low mass-galaxies ($10^9M_\odot$), the total time to merge the SMBHs exceeds $\sim 1$\,Gyr, whereas for high-mass galaxies ($10^{11}M_\odot$) the coalescence time is well
below $0.5$\,Gyr. 
This is contrary to the usual statement that binary coalescensce is slower in
more massive binaries.
That statement reflects the much longer timescales for gravitational encounters
in massive galaxies; but in our merger models, 
evolution prior to the GW regime occurs
much more rapidly than predicted based on collisional loss cone repopulation rates,
and the dominant influence on the coalescence time is the timescale for GW energy
loss, which is longer for smaller SMBH masses.
We note that all of the coalescence times in Table 3 are short compared with those
expected in spherical models (e.g. Figure 19 of Merritt, Mikkola \& Szell 2007).

The lower mass scaling corresponds to sources detectable by LISA.
We find coalescence times of a few Gyr in these cases, which, while
relatively short, are nevertheless long enough 
that SMBH binaries in these galaxies may 
not achieve full coalescence before a subsequent galaxy merger occurs.

\section{Summary}\label{sec-concl}

The ``final parsec problem'' refers to the difficulty of bringing binary
 supermassive black holes (SMBHs) to full coalescence,
due to rapid depletion, via slingshot ejection, of stars on 
orbits that intersect the massive binary.
It has been suggested \citep{MP2004} that binary hardening can occur much more efficiently
in triaxial galaxies due to the greater population of centrophilic orbits.
Generically, galaxies formed in dissipationless mergers are triaxial in shape,
and one might expect to avoid the final parsec problem in the case of
 binaries that form as a result of galaxy mergers.
To test this hypothesis,
we carried out high-accuracy, direct $N$-body simulations of dissipationless 
mergers containing SMBHs.
The results were compared with a second set of simulations of massive binaries 
in spherical models.
The strong $N$-dependence of the binary hardening rate in
the spherical geometry was found to vanish in the merger models.
This was true, even though the departures from axisymmetry in the merged galaxies
were slight.
The binary SMBHs in our merger simulations formed with large eccentricities, 
due to the merger geometry,
and eccentricities remained high during the subsequent 
evolution of the binary. 
Our results suggest that the ``final parsec problem'' may not be
as serious as previously believed, and that prompt coalescence of binary
SMBHs following mergers may be common, even in galaxies lacking gas.

\acknowledgments

This work began as a student project during the 
Advanced School and Workshop on Computational Gravitational Dynamics
held at the Lorentz Center, Leiden from 3 May 2010 - 13 May 2010.
We thank Simon Portegies Zwart for his efforts in organizing this workshop.
We thank Peter Berczik and Ingo Berentzen for their support on the numerical simulations and helpful discussions.
FK was supported by a grant from the Higher Education Commission (HEC) of 
Pakistan administrated by the Deutscher Akademischer Austauschdienst (DAAD).
DM was supported by grants AST-0807910 (NSF) and NNX07AH15G (NASA).
This research and the computer hardware used in Heidelberg
were supported by project ``GRACE'' I/80 041-043 of the
Volks\-wagen Foundation, by the Ministry of Science, Research
and the Arts of Baden-W\"urttemberg (Az: 823.219-439/30
and 823.219-439/36), and in part by the
German Science Foundation (DFG) under SFB 439 (sub-project
B11) "Galaxies in the Young Universe" at the University of
Heidelberg.
The $HPC-GPU$ cluster ``Kolob'' is supported by the University of Heidelberg through the excellence initiative for German universities by the innovation funds FRONTIER.

\begin{table*}
\caption{Time to Gravitational Wave Coalescence} 
\begin{tabular}{cccccccc }
\hline
Run & $a_\mathrm{final}$  & $s_\mathrm{final}$ & $e_0$ & $a_\mathrm{0}$ (pc) & $t_\mathrm{0}$ (Gyr) & $t_\mathrm{coal}$ (Gyr) \\
\hline
C1 & $1.2\times10^{-4}$ & $120.3$ & $0.51$ &  $(2.6\times10^{-3},1.2\times10^{-2})$ & $(11.5,0.54)$ & $(15.9,0.94)$   \\
C2 & $7.2\times10^{-5}$ & $115.2$  & $0.96$& $(1.2\times10^{-2},8.5\times10^{-2})$ 	& $(2.6,0.21)$ & $(5.1,0.29)$    \\
C3 & $7.3\times10^{-5}$ & $108.3$  & $0.80$& $(4.9\times10^{-3},3.3\times10^{-2})$ 	& $(7.5,0.35)$ & $(10.8,0.41)$    \\
C4 & $6.8\times10^{-5}$ & $118.2$  & $0.95$& $(1.0\times10^{-2},8.3\times10^{-2})$ 	& $(2.8,0.22)$ & $(5.7,0.30)$    \\
D1 & $5.2\times10^{-4}$ & $13.0$  & $0.97$& $(9.0\times10^{-2},6.7\times10^{-1})$ 	& $(2.7,0.21)$ & $(5.5,0.35)$    \\
D2 & $4.0\times10^{-4}$ & $11.2$  & $0.88$& $(4.4\times10^{-2},3.0\times10^{-1})$ 	& $(6.7,0.26)$ & $(10.8,0.47)$    \\
D3 & $5.1\times10^{-4}$ & $10.7$  & $0.95$& $(7.2\times10^{-2},5.4\times10^{-1})$ 	& $(4.0,0.22)$ & $(7.6,0.38)$    \\
D4 & $4.7\times10^{-4}$ & $11.0$  & $0.98$& $(9.0\times10^{-2},9.1\times10^{-1})$ 	& $(2.2,0.23)$ & $(4.7,0.26)$    \\
\hline
\end{tabular}\label{GW}
\end{table*}

\begin{table*}
\caption{Time to Gravitational Wave Coalescence (Run C3)} 
\begin{tabular}{cccccccc }
\hline
$e_0$ & $a_\mathrm{0}$ (pc) & $t_\mathrm{coal}$ (Gyr) \\
\hline
$0.5$ & $(2.6\times10^{-3},1.8\times10^{-2})$ & $(14.96,0.65)$   \\
$0.0$ & $(1.9\times10^{-3},1.3\times10^{-2})$ & $(20.46,0.86)$   \\
\hline
\end{tabular}\label{TableGWC3}
\end{table*}

\end{document}